\documentclass[12pt]{iopart}
\usepackage{graphicx}
\usepackage{color}
\begin{document}

\title[]{
Charge order and possible bias-induced metastable state in the organic conductor 
$\beta$-({\it meso}-DMBEDT-TTF)$_2$PF$_6$: effects of structural distortion}

\author{Y Tanaka and K Yonemitsu}

\address{Department of Physics, Chuo University, Bunkyo-ku, Tokyo 112-8551, Japan}
\address{JST, CREST, Chiyoda-ku, Tokyo 102-0076, Japan}
\ead{yasuhiro@phys.chuo-u.ac.jp}      

\begin{abstract}
We theoretically investigate charge order and nonlinear conduction 
in a quasi-two-dimensional organic conductor $\beta$-({\it meso}-DMBEDT-TTF)$_2$PF$_6$ 
[DMBEDT-TTF=dimethylbis(ethylenedithio)tetrathiafulvalene].
Within the Hartree-Fock approximation, we study effects of structural distortion on the 
experimentally observed checkerboard charge order and its bias-induced melting by using 
an extended Hubbard model with Peierls- and Holstein-types of electron-lattice interactions.
The structural distortion is important in realizing the charge order. 
The current-voltage characteristics obtained by a nonequilibrium 
Green's function method indicate that a charge-ordered insulating state changes into a 
conductive state. Although the charge order and lattice distortions are 
largely suppressed at a threshold voltage, they remain finite even in the conductive state. 
We discuss the relevance of the results to experimental observations, especially to a 
possible bias-induced metastable state.
\end{abstract}



\maketitle
\clearpage

\section{Introduction}
Organic conductors offer intriguing subjects of low-dimensional correlated 
electron systems in which various electronic states emerge at low 
temperatures\cite{Ishiguro,Seo_ChemRev04}. In addition to the origin of such 
phases in thermal equilibrium condition, nonequilibrium properties induced 
by external stimuli have attracted much attention 
recently\cite{Yonemtsu_PhysRep08,Mori_JMC07}. For example, photoinduced phase 
transitions\cite{Iwai_PRL07,Onda_PRL08,Kawakami_PRL09} and nonlinear conduction 
phenomena\cite{Kumai_SCI99,Sawano_NATURE05,Takahide_PRL06,Mori_PRL08}, which have been 
observed mainly in Mott insulators and charge-ordered compounds, have been 
intensively studied.

In the quasi-two-dimensional organic conductor 
$\beta$-({\it meso}-DMBEDT-TTF)$_2$PF$_6$\cite{Kimura_Chcm04} [abbreviated as 
$\beta$-(DMeET)$_2$PF$_6$ hereafter], nonlinear conduction has been observed 
in a charge-ordered insulating phase\cite{Niizeki_JPSJ08}. The conduction 
layer of $\beta$-(DMeET)$_2$PF$_6$ consists of a weakly dimerized pair of 
neighboring donor molecules [Fig. 1(a)] and forms a 3/4-filled band. 
This compound shows a second-order metal-insulator transition at 90 K, 
where the peculiar checkerboard charge 
order (CCO) occurs\cite{Kimura_JACS06,Mori_JPSJ06} [Fig. 1(b)]. This transition 
accompanies a structural distortion. Under pressure, the CCO is readily 
suppressed and superconductivity with $T_C\sim 4$ K 
appears\cite{Kimura_Chcm04,MTanaka_JPSJ08,Morinaka_PRB09}.

Theoretically, Yoshimi {\it et al.}\cite{Yoshimi_JPSJ07} have investigated the 
superconductivity by 
using the random phase approximation for an extended Hubbard model with on-site 
and nearest neighbor Coulomb interactions. They show that the CCO does not 
appear when they consider 
values of the Coulomb interactions that are determined from the intermolecular distances 
and the effective molecular sizes. Since their model has only electronic degrees of freedom, 
their result suggests 
that electron-lattice (e-l) interactions play an important role in stabilizing the CCO. 
In fact, it is experimentally known that modulation of transfer integrals and 
bending of molecules occur at the transition triggered by the charge 
disproportionation\cite{Kimura_JACS06}. However, the effects of lattice degrees of 
freedom on the CCO have not been studied so far. 

When the bias voltage is applied, $\beta$-(DMeET)$_2$PF$_6$ exhibits 
a negative differential resistance below 70 K, where the 
initial CCO insulating state shows a steep decrease of the 
resistivity\cite{Niizeki_JPSJ08}. 
Moreover, in the current-voltage characteristics, a non-monotonic 
behavior of the voltage drop was observed. 
A recent Raman spectroscopy measurement\cite{Niizeki_PhysB10} 
indicates that the CCO melts by the bias. The time-resolved measurement shows 
a two-stepped sample voltage drop with a transient plateau, which suggests 
that there appears an 
electric-field-induced metastable state which is different 
from the normal metallic phase at high 
temperatures\cite{Niizeki_JPSJ08,Niizeki_PhysB10}. 
Such a new phase in nonequilibrium condition is also suggested in a 
photoinduced state of a quasi-one-dimensional organic conductor 
(EDO-TTF)$_2$PF$_6$ in which strong electron-phonon interactions play an important 
role in its charge-ordered ground state\cite{Onda_PRL08,Clay_PRB03,Yonemitsu_PRB07,Gao_Nature13}. 
In a recent femtosecond electron diffraction study\cite{Gao_Nature13}, 
time-dependent structural changes after 
the photoexcitation have been monitored and the difference between the photoinduced transient 
state and the low-temperature charge-ordered state has been identified. 
In this context, in addition to the origin of the CCO in $\beta$-(DMeET)$_2$PF$_6$, 
the mechanism of the nonlinear conduction and the identification of the phases 
appearing by melting the CCO are intriguing subjects. 

In this paper, we study the equilibrium and nonequilibrium properties of 
$\beta$-(DMeET)$_2$PF$_6$. We use an extended Hubbard model with two 
kinds of e-l couplings that are deduced from the experimental observations. 
The CCO becomes the ground state by these lattice distortions since the Coulomb 
interactions basically do not favor the checkerboard charge pattern. 
This state is similar to the (1100) charge order that has been discussed 
in quasi-one-dimensional systems by Clay {\it et al.}\cite{Clay_PRB03}. 
In two dimensions, a concept of the paired-electron crystal\cite{Li_JPCM10,Dayal_PRB11} 
that consists of pairs of charge-rich 
sites separated by pairs of charge-poor sites, has been proposed in order to 
understand the ground state of interacting quarter-filled systems 
with e-l couplings and frustration. 
We investigate nonlinear conduction by applying a bias to the CCO insulating state. 
The bias induces a conductive state possessing a weak CCO and partial 
lattice distortions, which is distinct from the insulating ground state. 
This indicates the appearance of a metastable state different from the 
metallic phase at high temperatures. 

This paper is organized as follows. In \S 2, the model Hamiltonian for 
$\beta$-(DMeET)$_2$PF$_6$ and the methods of calculations are presented. In \S 3, we 
show the results and their relevance to the experimental observations is discussed. 
\S 4 is devoted for summary and conclusions.

\section{Model and Methods}
We consider the extended Hubbard model with Peierls- and Holstein-types of 
e-l interactions written as
\begin{eqnarray}
H&=&\sum_{\langle ij \rangle \sigma} \left[
(t_{i,j} + \alpha u_{i,j}) c^\dagger_{i\sigma} c_{j\sigma} +h.c.
\right] \nonumber \\ 
&+&U\sum_i (n_{i\uparrow}-N_e/2)(n_{i\downarrow}-N_e/2)\nonumber \\
&+&\sum_{\langle ij \rangle} V_{i,j}(n_i-N_e)(n_j-N_e)+\beta\sum_{i}v_i(n_i-N_e)\nonumber \\
&+&\sum_{\langle ij \rangle } \frac{K_{\alpha}}{2} u_{i,j}^2
+\sum_{i} \frac{K_{\beta}}{2} v_{i}^2, \
\label{eq:eq1}
\end{eqnarray}
where $\langle ij\rangle$ represents a pair of neighboring sites, 
$c^{\dagger}_{i\sigma}(c_{i\sigma})$ denotes the
creation (annihilation) operator for an electron with spin $\sigma$ at
the $i$th site, $n_{i\sigma}=c^{\dagger}_{i\sigma}c_{i\sigma}$, 
$n_{i}=n_{i\uparrow}+n_{i\downarrow}$, and the average electron density
is $N_e=1.5$. We write the transfer integrals as $t_{i,j}$, the on-site repulsion $U$, 
and the nearest neighbor Coulomb interactions $V_{i,j}$.
The coupling strength, the lattice displacement, and the elastic constant
for the Peierls-type e-l interaction are denoted by $\alpha$, $u_{i,j}$, 
and $K_{\alpha}$, respectively, whereas those for the Hostein-type e-l interaction 
are written as $\beta$, $v_{i}$, and $K_{\beta}$. We introduce new variables by
\begin{equation}
y^{i,j}_{\alpha}=\alpha u_{i,j},\nonumber
\end{equation}
\begin{equation}
s_{\alpha}=\alpha^{2}/K_{\alpha},\nonumber
\end{equation}
\begin{equation}
y^{i}_{\beta}=\beta v_{i},\nonumber
\end{equation}
\begin{equation}
s_{\beta}=\beta^{2}/K_{\beta}.\nonumber
\end{equation}

For simplicity, we consider only 
one mode of displacements for each type of e-l interactions. The definition of the 
actual modulations of transfer integrals and on-site potentials is given below.
\begin{figure}
\includegraphics[height=7.5cm]{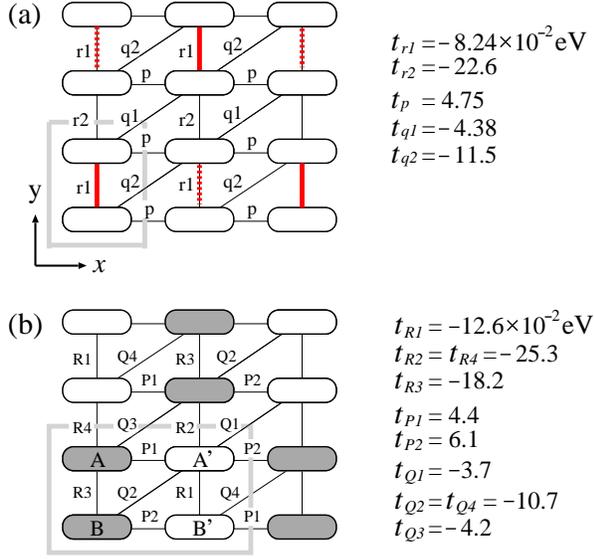}
\caption{(Color online) Schematic representation of molecular configurations in 
donor layer of $\beta$-(DMeET)$_2$PF$_6$: (a) high-temperature metallic 
phase and (b) CCO phase. 
The gray rectangles indicate the unit cells. The transfer integrals estimated from the 
extended H$\ddot{\rm u}$ckel method\cite{Kimura_Chcm04,Mori_JPSJ06} are also shown. 
In (a), the red or thick solid (dashed) lines indicate the bonds on which the 
magnitudes of the transfer integrals are increased (decreased) by the distortions 
considered in the present paper. In (b), the sites $A$ and $B$ are hole-rich, 
whereas the sites $A^{\prime}$ and $B^{\prime}$ are hole-poor.}
\label{fig:fig1}
\end{figure}

A schematic illustration of the high-temperature metallic phase of
$\beta$-(DMeET)$_2$PF$_6$ is shown in Fig. \ref{fig:fig1}(a). 
The unit cell contains two molecules. The charge density per molecule 
is uniform in this phase. The transfer integrals $t_{r1}$ and $t_{r2}$ 
are alternating on the vertical direction. The weakly dimerized molecules 
are connected by $r2$ bonds on which the magnitude of the transfer integrals 
is the largest. 

In the CCO phase, there are four molecules in the unit cell as shown in 
Fig. \ref{fig:fig1}(b). The transfer integrals are modulated by the structural 
distortion. The modulations on the vertical bonds are larger than those 
on the diagonal and horizontal bonds. In particular, $r1$ bonds in the high-temperature 
phase change into $R1$ and $R3$ bonds that are inequivalent ($t_{R1}>t_{R3}$). 
According to the X-ray diffraction measurement\cite{Kimura_JACS06,Mori_JPSJ06}, 
we have $d_{R2}=d_{R4}<d_{R3}\simeq d_{R1}$, where $d_{l}$ denotes the 
distance between the neighboring molecules connected by the bond ${l}$. 
The intermolecular distance within a dimer is the shortest.
For the diagonal and horizontal directions, the intermolecular distances are 
larger than $d_{R1}$ and $d_{R3}$. With these facts, we expect that 
$V_{r2}>V_{r1}>V_{p},\ V_{q1},\ V_{q2}$ in Fig. \ref{fig:fig1}(a), where $V_{l}$ represents 
the nearest neighbor Coulomb interaction between the two sites connected by the bond $l$. 
In fact, it has been pointed 
out\cite{Kimura_JACS06,Mori_JPSJ06} that, if it were not for e-l interactions, 
the charge order along the 
vertical direction should be of r-p-r-p type, where r and p represent hole-rich 
and -poor sites, respectively. This is because the charge pattern of the 
intradimer pair should be r-p, and that of the second-nearest interdimer pair 
should also be r-p if we consider only the Coulomb interactions that are 
determined by the intermolecular distances. However, the CCO is of r-r-p-p type 
in which the two molecules connected by $R3$ ($R1$) bonds become hole-rich 
(hole-poor). These considerations suggest that lattice effects are the key to 
understanding the CCO. 

In order to take account of the structural distortion, we introduce modulations 
of the transfer integrals on $r1$ bonds as 
\begin{equation}
t^{d}_{R1}=t_{r1}+y_{\alpha},
\end{equation}
\begin{equation}
t^{d}_{R3}=t_{r1}-y_{\alpha},
\end{equation}
where $t^{d}_{R1}$ ($t^{d}_{R3}$) is the modified transfer integral located 
on $R1$ ($R3$) bonds in the low-temperature structure [Fig. \ref{fig:fig1}(b)]. 
When $y_{\alpha}>0$, we have $t^{d}_{R1}>t^{d}_{R3}$ that 
is consistent with the relation $t_{R1}>t_{R3}$ in Fig. \ref{fig:fig1}(b), which
is derived from the experimental result. Experimentally, the thermal 
contraction makes the averaged $|t|$ larger at low temperatures, thus 
$|t_{R1}|, |t_{R3}|>|t_{r1}|$. 
In essence, the modulation given by Eqs. (6) and (7) stabilizes the r-r-p-p type charge 
pattern in the CCO. We do not consider modulations of the other transfer 
integrals since they are relatively small compared to the above. 
In our calculations, the values of the transfer 
integrals $t_{i,j}$ in Eq. (1) are fixed at those in the 
high-temperature phase. 
We use $t_{r1}=-0.0824$, $t_{r2}=-0.226$, $t_{p}=0.0475$, 
$t_{q1}=-0.0438$, and $t_{q2}=-0.115$. 

For the Holstein-type e-l coupling, we assume
\begin{eqnarray}
y^{i}_{\beta}=\left\{
\begin{array}{ll}
y_{\beta}&\ {\rm for}\ i\in A,B\\
-y_{\beta}&\ {\rm for}\ i\in  A^{\prime},B^{\prime}.\\
\end{array}\right.
\end{eqnarray}
Experimentally, it has been known that in the CCO phase the hole-poor molecules 
are slightly bent, whereas the hole-rich molecules are almost 
flat\cite{Kimura_JACS06}. This indicates that the modulation of the on-site 
potential is accompanied by the CCO.

\begin{figure}
\includegraphics[height=4.5cm]{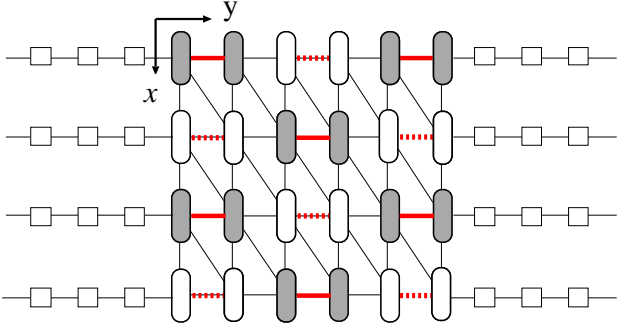}
\caption{(Color online) Schematic model for nonequilibrium properties. 
The left and right electrodes are attached to the central part where the CCO is realized 
when the bias is absent. The $y$-axis ($x$-axis) is along (perpendicular to) the 
conduction direction.}
\label{fig:fig2}
\end{figure}

We apply the Hartree-Fock approximation to the electron-electron interactions 
in Eq. (\ref{eq:eq1}). In the equilibrium case, we calculate 
the ground-state energies for several kinds of possible charge-order patterns 
with 2$\times$4 or 4$\times$2 unit cell and the periodic boundary condition for 
much larger systems. The stability 
of the CCO due to the e-l couplings is examined. 
The lattice distortions $y_{\alpha}$ and $y_{\beta}$ are self-consistently determined 
by the Hellmann-Feynman 
theorem. In the nonequilibrium case, we calculate physical quantities such as the 
current-voltage ($J$-$V$) characteristics by using the nonequilibrium Green's 
function method\cite{Yonemitsu_JPSJ09,Tanaka_PRB11} which treats nonequilibrium 
steady states and contains the definitions of $J$ and $V$. This 
method has been used to study the suppression of rectification\cite{Yonemitsu_JPSJ09} 
and the dielectric breakdown\cite{Tanaka_PRB11} in one-dimensional interacting 
electron systems. The results are basically consistent with those obtained by 
other numerical approaches which take account of the effects of 
quantum fluctuations\cite{Yonemitsu_PRB07_2,Heidrich_PRB10}. Although the Hartree-Fock 
approximation generally overestimates the stability of ordered states, the present 
approach is expected to capture the essential physics of nonequilibrium phenomena 
in $\beta$-(DMeET)$_2$PF$_6$. 
As shown in Fig. \ref{fig:fig2}, we attach left and right 
semi-infinite metallic electrodes to the CCO system that is referred to as the 
central part. 
We choose the $y$-axis in Fig. 1(a) as the conduction direction and the qualitative 
results are unaltered if the bias is applied along the $x$-axis. 
We assume that the electrodes are 
one-dimensional and the electrons in them are noninteracting. We do not 
consider work-function difference at the interfaces for simplicity. 
For the electrodes, the wide-band limit is applied so that the retarded self-energies 
due to the electron transfers between the electrodes and the central part are independent 
of energy\cite{Jauho_PRB94}. For finite $V$, we introduce the chemical 
potential $\mu_C=(\mu_L+\mu_R)/2$\cite{Yonemitsu_JPSJ09} and adjust it such that the 
electron density of the central part is fixed at 3/4-filling. 
Here $\mu_L=\mu_C+V/2$ and $\mu_R=\mu_C-V/2$ are the left 
and right chemical potentials, respectively. Throughout the paper, 
we set $\gamma_{L}=\gamma_{R}=0.1$ where $\gamma_{L}$ ($\gamma_{R}$) is the coupling 
constant between the central part and the left (right) 
electrode\cite{Yonemitsu_JPSJ09,Tanaka_PRB11}, and $e=\hbar=1$. The interaction 
parameters $U$, $V_{i,j}$, $s_{\alpha}$, and $s_{\beta}$ are given in units of eV in the 
following.

\section{Results}
\begin{figure}
\includegraphics[height=5.5cm]{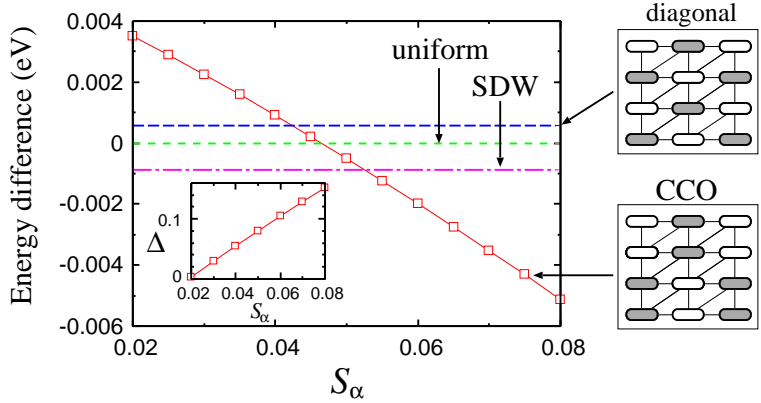}
\caption{(Color online) Ground-state energy per site of CCO as a function of $s_{\alpha}$ 
for $U=0.5$, $V_{r1}=0.23$, $V_{r2}=0.26$, $V_{p}=V_{q1}=V_{q2}=0.2$, $s_{\beta}=0.45$, 
and $L_x=L_y=32$. The energies per site of the paramagnetic metallic phase (uniform), 
the diagonal CO, and the spin-density-wave state (SDW), which are obtained without 
the e-l couplings, are also shown. The energy of the uniform state is chosen to zero. 
The inset shows the charge gap $\Delta$ in the CCO state as a function of $s_{\alpha}$.}
\label{fig:fig3}
\end{figure}

\begin{figure}
\includegraphics[height=5.5cm]{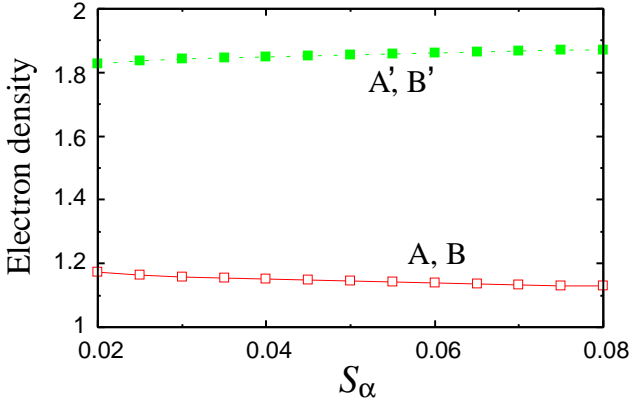}
\caption{(Color online) Electron density of CCO phase as a function of $s_{\alpha}$. 
The other parameters are the same as in Fig. \ref{fig:fig3}.}
\label{fig:fig4}
\end{figure}

\begin{figure}
\includegraphics[height=5.5cm]{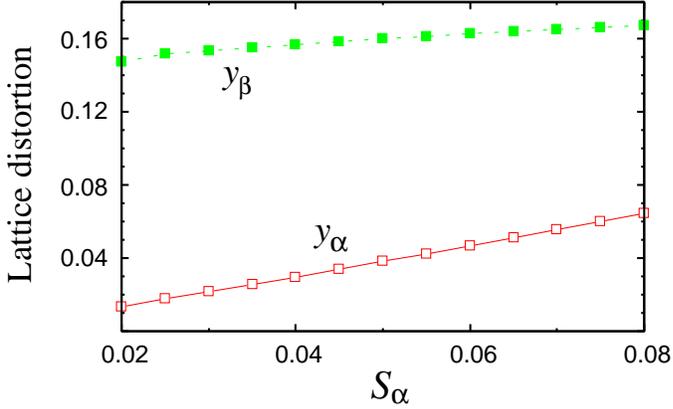}
\caption{(Color online) Lattice distortions $y_{\alpha}$ and $y_{\beta}$ as a function 
of $s_{\alpha}$. The other parameters are the same as in Fig. \ref{fig:fig3}.}
\label{fig:fig5}
\end{figure}

First, we discuss the ground state without the bias voltage. 
For the electron-electron interactions, we use 
$U=0.5$, $V_{r1}=0.23$, $V_{r2}=0.26$, and $V_{p}=V_{q1}=V_{q2}=0.2$. 
These parameter values are comparable to those obtained in recent 
{\it ab initio} calculations with the screening effects for typical organic 
compounds\cite{Nakamura_PRB12}. For $V_{i,j}$, we choose the values so that they satisfy 
the relation $V_{r2}>V_{r1}>V_{p},\ V_{q1},\ V_{q2}$ as discussed in \S 2. 
In Fig. \ref{fig:fig3}, we show the ground-state energy of the CCO for 
$s_{\beta}=0.45$ as a function of $s_{\alpha}$. The system size is $L_x=L_y=32$ where 
$L_x$ ($L_y$) is the number of sites in the $x$-direction ($y$-direction). 
In this figure, we also show the energies of three other phases, a paramagnetic metallic state 
with uniform charge density, a spin-density-wave (SDW) state and a diagonal CO, 
which are obtained without the e-l
couplings. We have confirmed that the lattice distortions in Eqs. (6)-(8) 
do not lower the energies of the SDW and the diagonal CO. 
The uniform phase is taken as a reference state and its energy is chosen to be zero. 
The diagonal CO has a slightly higher energy than the uniform phase. 
In the SDW state, the spins are antiferromagnetic along the $x$-axis, whereas they are
ferromagnetic along the $y$-axis. This is consistent with the results in the random phase
approximation\cite{Yoshimi_JPSJ07}, which show that the spin susceptibility has the 
largest peak at ${\bf q}=(\pi, 0)$ for $U=0.5$. 
When the e-l couplings are present, the CCO becomes the ground state for 
$s_{\alpha}>0.055$. This state has a charge gap $\Delta$ as shown in the inset of 
Fig. \ref{fig:fig3}. In Figs. \ref{fig:fig4} and \ref{fig:fig5}, we show the 
electron density and the lattice distortions $y_{\alpha}$ and $y_{\beta}$. The electron 
density on sites $A$ ($A^{\prime}$) is the same as that on sites $B$ ($B^{\prime}$). 
The CCO in Fig. \ref{fig:fig3} is nonmagnetic. In fact, a CCO with 
an antiferromagnetic order on the hole-rich sites, which has a slightly lower energy than 
the nonmagnetic CCO, is also obtained as a mean-field solution. The energy 
difference between the two CCOs is smaller than 10$^{-3}$ eV/site for $s_{\alpha}=0.06$ and the 
difference decreases with increasing $s_{\alpha}$. Because the antiferromagnetic order is an 
artifact of the Hartree-Fock approximation, 
hereafter we use the nonmagnetic CCO as the ground state when the bias 
voltage is applied. It should be noted that the spin configuration does not qualitatively 
alter the results. According to the experimental observations, the magnetic susceptibility 
for $\beta$-(DMeET)$_2$PF$_6$ shows gradual decrease with lowering temperatures and abruptly 
drops around $T=80$ K\cite{Shikama_CRYSTAL12}. Its temperature dependence below 80 K suggests 
that the system is nonmagnetic and a spin gap opens in the CCO state.


\begin{figure}
\includegraphics[height=5.5cm]{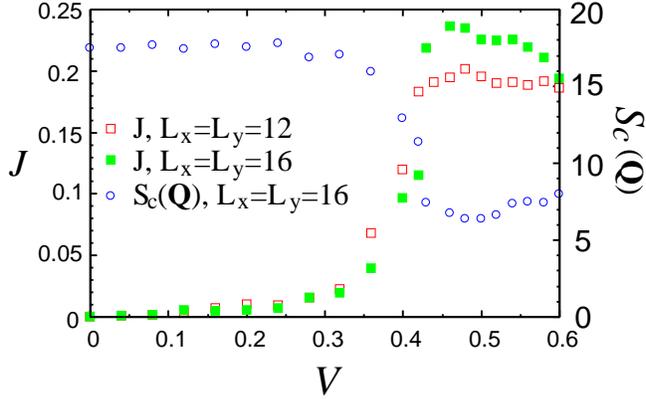}
\caption{(Color online) $J$-$V$ characteristics and $V$ dependence of charge 
structure factor $S_c({\bf Q})$ for $L_x=L_y=16$ with $s_{\alpha}=0.075$. $J$-$V$ curve 
for $L_x=L_y=12$ is also shown. The other parameters are the same as those in Fig. 
\ref{fig:fig3}.}
\label{fig:fig6}
\end{figure}

\begin{figure}
\includegraphics[height=5.5cm]{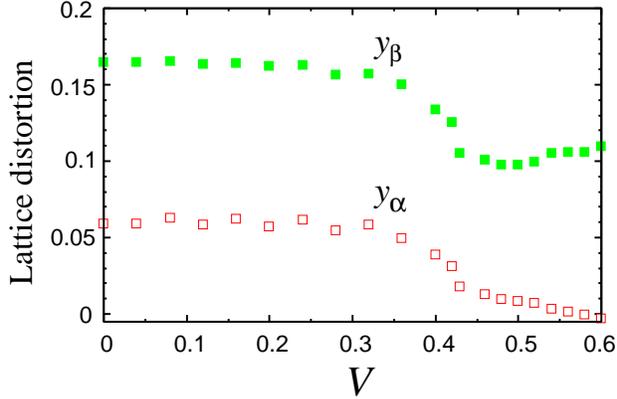}
\caption{(Color online) Lattice distortions $y_{\alpha}$ and $y_{\beta}$ 
as a function of $V$. The parameters are the same as those in Fig. \ref{fig:fig6}.}
\label{fig:fig7}
\end{figure}

\begin{figure}
\includegraphics[height=5.5cm]{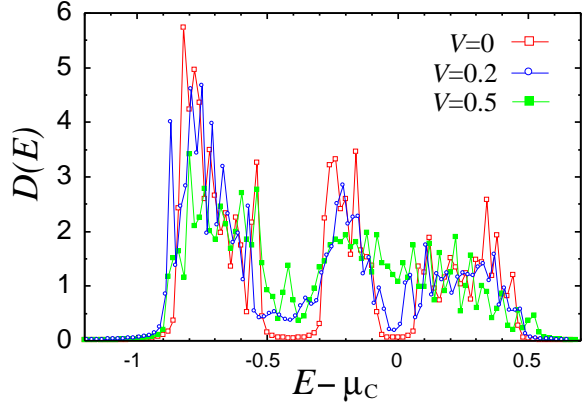}
\caption{(Color online) Density of states for $V=0$, $0.2$, and $0.5$. 
The parameters are the same as those in Fig. \ref{fig:fig6}.}
\label{fig:fig8}
\end{figure}

For finite bias voltage $V$, we set $s_{\alpha}=0.075$ so that the CCO becomes 
the ground state for $V=0$. The effects of the bias on the CCO are investigated by the 
charge structure factor $S_c({\bf q})$, which is defined as 
\begin{equation}
S_c({\bf q})=\frac{1}{N_s}\sum_{\mu,\nu}(\langle n_{\mu ,1}n_{\nu ,1}\rangle
+\langle n_{\mu ,2}n_{\nu ,2}\rangle )e^{i{\bf q}({\bf R}_{\mu}-{\bf R}_{\nu})},
\end{equation}
where $\mu$ and $\nu$ denote indices for the unit cell shown in Fig. \ref{fig:fig1}(a), 
1 and 2 represent indices for the sites inside the unit cell, and ${\bf R}_{\mu}$ (${\bf R}_{\nu}$) 
is the position vector of the $\mu$-th ($\nu$-th) unit cell. $N_s=L_xL_y$ is the total number 
of sites in the central part. The wave vector that corresponds 
to the CCO is ${\bf q}={\bf Q}=(\pi, \pi)$. In Fig. \ref{fig:fig6}, we show the $J$-$V$ 
characteristics and the $V$ dependence of $S_c({\bf Q})$ for $L_x=L_y=16$. For comparison, 
we show the $J$-$V$ curve obtained with $L_x=L_y=12$, which gives qualitatively 
the same results. For small $V$, the current $J$ does not flow because the charge gap opens 
at $V=0$ owing to the CCO. In this region, $S_c({\bf Q})$ is almost unchanged, so the CCO is 
robust against the bias. With increasing $V$, $J$ abruptly increases at 
$V=V_{\rm th}\sim 0.4$, where the CCO changes into a conductive state. 
Although $S_c({\bf Q})$ largely decreases around $V=V_{th}$, 
it remains finite even in $V>V_{th}$, which means that 
the CCO is not completely destroyed in the conductive state. 
In Fig. \ref{fig:fig7}, we show the lattice distortions $y_{\alpha}$ and $y_{\beta}$ 
as a function of $V$. 
Similar to the $V$ dependence of $S_c({\bf Q})$, $y_{\alpha}$ and $y_{\beta}$ do not show 
noticeable change for $V<V_{th}$. At $V=V_{th}$, both distortions are suppressed. 
In particular, the $y_{\alpha}$-distortion almost disappears for $V>V_{th}$, whereas 
the $y_{\beta}$-distortion survives. 
In equilibrium conditions, the CCO has a finite gap $\Delta$ as shown in Fig. \ref{fig:fig3}. 
However, $\Delta$ becomes very small when $y_{\alpha}$ decreases. When the bias is applied, 
the initial insulating CCO switches to the conductive state by the suppression of the 
$y_{\alpha}$-distortion. The weak CCO survives owing to the partial 
$y_{\beta}$-distortion, so that the bias-induced state is different from the uniform phase or 
the CCO at $V=0$. In Fig. \ref{fig:fig8}, we show the density of states 
$D(E)$ for different values of $V$. For $V=0$, there is no state around $E=\mu_C$ because 
the charge gap opens. For $V=0.2$, the gap structure in $D(E)$ still exists since the CCO 
and the lattice distortions at $V=0$ are robust against the bias for small $V$. 
When $V=0.5>V_{th}$, the gap at $E=\mu_C$ disappears, which indicates that the weak CCO 
has the conduction property. 
In Fig. \ref{fig:fig6}, the current $J$ begins to decrease when $V$ approaches half the 
bandwidth $W=1.29$. This is because the number of one-particle states that have large 
contributions to the current $J$ decreases for $V>W/2$\cite{Tanaka_PRB11}. In other words, 
the tilting of the band by the applied bias becomes large, so that the ballistic transport 
is suppressed. Effects of the inelastic scattering in the central 
part, which are not considered in the present calculations, will modify the $J$-$V$ 
curve in such a large $V$ region. 


Our results indicate the appearance of a bias-induced metastable state accompanied by a 
weak CCO. This state has only the Holstein-type distortion, which is weakened compared 
to the case with $V=0$. 
Experimentally, the time-resolved voltage measurement\cite{Niizeki_JPSJ08} 
has shown that a two-stepped sample voltage drop with a transient plateau 
occurs below 70 K, which indicates that a metastable state actually exists. 
Such a two-stepped structure is also observed in the $J$-$V$ 
characteristics\cite{Niizeki_JPSJ08}. 
According to the Raman spectroscopy measurement\cite{Niizeki_PhysB10}, 
in this transient metastable state the modes that are characteristic of the 
charge disproportionation almost disappear, so that the metastable state is ascribed to
the charge fluctuations. 
At present, it is difficult to compare our results directly to the experimental 
observations. Effects of the charge fluctuations that are ignored in the Hartree-Fock 
approximation may be important. 
The lattice modulation of $\beta$-(DMeET)$_2$PF$_6$ may be more complex than that 
considered in the present paper. 
In any case, our results suggest that the weak CCO has only the Holstein-type 
distortion in the metastable state. If we calculate the electronic state for $V=0$ by 
using values of the Peierls-type and Holstein-type distortions in the metastable state, 
the resulting CCO has a metallic band structure. 
This originates from the suppression of the Peierls-type distortion. 
In this state, the values of $D(E)$ around $E=\mu_C$ are comparable 
to those at $V=0.5$ in Fig. \ref{fig:fig8}. 
The applied bias converts the insulating CCO into the conductive CCO. 
Our results may be related to the photoinduced structural changes in 
(EDO-TTF)$_2$PF$_6$ which exhibits the (1100) charge order at low temperatures. This 
charge order is triggered by displacements and bending of the EDO-TTF molecules, which 
corresponds to the Peierls- and Holstein-types of e-l couplings, 
respectively\cite{Clay_PRB03,Yonemitsu_PRB07}. 
The femtosecond electron diffraction study for (EDO-TTF)$_2$PF$_6$ has shown that 
the photoexcitation induces the metastable state mainly through the suppression of the 
molecular displacements, whereas the molecular bendings do not show noticeable 
change\cite{Gao_Nature13}. 
More structural information such as X-ray diffraction data for the present material in 
the presence of the bias voltage may clarify the nature of this novel nonequilibrium phase. 

\section{Summary}
We investigate effects of the e-l couplings on the equilibrium and nonequilibrium 
properties of $\beta$-(DMeET)$_2$PF$_6$. We have shown that the CCO that is 
peculiar to this compound is obtained by the extended Hubbard model with two 
kinds of e-l interactions that originate from molecular displacements and 
deformations. The e-l couplings are important in stabilizing the CCO. 
The applied bias changes the initial insulating CCO into a 
conductive state with the weak CCO that emerges by the disappearance of the 
Peierls-type distortion. 
This indicates the presence of a bias-induced metastable state different from the 
high-temperature metallic state. These results suggest that the lattice 
degrees of freedom play a key role in determining the conduction property in 
$\beta$-(DMeET)$_2$PF$_6$. 

\ack
This work was supported by a Grant-in-Aid for Young Scientists (B) (Grant No. 
12019365) from the Ministry of Education, Culture, Sports, Science and 
Technology of Japan.

\end{document}